\documentclass[twocolumn,aps,prl,showpacs, superscriptaddress, amsmath,amssymb]{revtex4}

\usepackage{graphicx}
\usepackage{dcolumn}
\usepackage{bm}
\usepackage{epstopdf}
\usepackage{color}
\usepackage{pdfpages}

\begin{document}

\title{Diffusion of ellipsoids in bacterial suspensions}
\author{Yi Peng}
\affiliation{Department of Chemical Engineering and Materials
Science, University of Minnesota, Minneapolis, MN 55455, USA}
\author{Lipeng Lai}
\affiliation{Beijing Computational Science Research Center, Beijing 100193, China}
\author{Yi-Shu Tai}
\affiliation{Department of Chemical Engineering and Materials
Science, University of Minnesota, Minneapolis, MN 55455, USA}
\author{Kechun Zhang}
\affiliation{Department of Chemical Engineering and Materials
Science, University of Minnesota, Minneapolis, MN 55455, USA}
\author{Xinliang Xu}
\email{xinliang@csrc.ac.cn}
\affiliation{Beijing Computational Science Research Center, Beijing 100193, China}
\author{Xiang Cheng}
\email{xcheng@umn.edu}
\affiliation{Department of Chemical Engineering and Materials
Science, University of Minnesota, Minneapolis, MN 55455, USA}

\date{\today}
\pacs{87.17Jj, 05.40-a, 47.63.Gd} \keywords{active matter, bacterial suspensions, Brownian diffusion}

\begin{abstract}

Active fluids such as swarming bacteria and motile colloids exhibit exotic properties different from conventional equilibrium materials. As a peculiar example, a spherical tracer immersed inside active fluids shows an enhanced translational diffusion, orders of magnitude stronger than its intrinsic Brownian motion. Here, rather than spherical tracers, we investigate the diffusion of isolated ellipsoids in a quasi-two-dimensional bacterial bath. Our study shows a nonlinear enhancement of both translational and rotational diffusions of ellipsoids. More importantly, we uncover an anomalous coupling between particles' translation and rotation that is strictly prohibited in Brownian diffusion. The coupling reveals a counterintuitive anisotropic particle diffusion, where an ellipsoid diffuses fastest along its minor axis in its body frame. Combining experiments with theoretical modeling, we show that such an anomalous diffusive behavior arises from the generic straining flow of swimming bacteria. Our work illustrates an unexpected feature of active fluids and deepens our understanding of transport processes in microbiological systems.                                    

\end{abstract}

\maketitle

The diffusion of a small tracer in a surrounding medium provides a reliable means for probing material properties of complex fluids \cite{Squires10}. Particularly, such a method has been used to investigate unique features of active fluids---a novel class of nonequilibrium soft materials with examples across a wide range of biological and physical systems including flocking animals \cite{Marchetti13,Koch11,Poon13}, vibrated granular beds \cite{Narayan07}, synthetic colloidal swimmers \cite{Palacci10,Palacci13}, and a self-propelled cytoskeleton \cite{Schaller10,Sanchez12}. The behavior of spherical tracers in active fluids is most clearly illustrated by the diffusion of colloidal spheres in suspensions of swimming microorganisms \cite{Wu00,Chen07,Wilson11,Mino13,Jepson13,Valeriani11,Leptos09,Kurtuldu11,Kasyap14}. Because of the hydrodynamic and steric interactions \cite{Underhill08,Ishikawa10,Lin11,Pushkin13,Morozov14}, a colloidal particle immersed in a bath of microswimmers exhibits a super-diffusive behavior at short times and a dramatically enhanced translational diffusion at long times. 

\begin{figure}
\begin{center}
\includegraphics[width=3.35in]{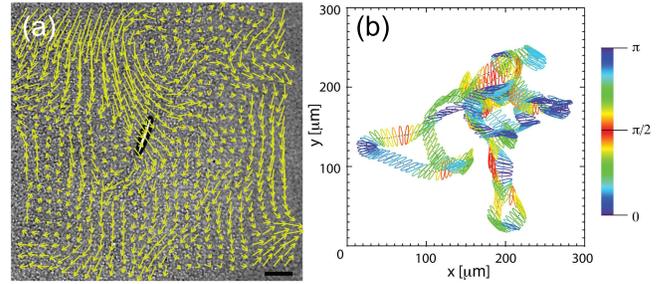}
\end{center}
\caption[Confined colloidal liquids in spherical cavities]{(Color online). Diffusion of an ellipsoid in quasi-two-dimensional bacterial bath. (a) Velocity field of swarming bacteria around an ellipsoid obtained from particle image velocimetry (PIV). Scale bar: 20 $\mu$m. (b) Trajectory of an ellipsoid in a time interval $\Delta t = 0.2$ s. The trajectory is obtained from a custom algorithm for particle tracking velocimetry (PTV). The center of mass of the particle at different times, $\bm{x}(t)$, is indicated by black dots. The color indicates the orientation of the particle, $\theta(t)$, with respect to the $x$ axis fixed in the lab frame. Bacterial concentration $n = 30n_0$.} \label{Figure1}
\end{figure}

The enhanced diffusion of passive particles such as nutrient granules, dead bacterial bodies and extracellular products is of great biological importance, which maintains an active ecological balance \cite{Wu00}, stimulates bio-mixing \cite{Kurtuldu11}, and promotes intercellular signaling and metabolite transports \cite{Morozov14}. However, few natural particles have the perfect spherical symmetry and usually possess more than translational degrees of freedom. It is still an open question how and to what extent the enhanced translational diffusion of an anisotropic particle is influenced by other degrees of freedom especially by its rotation. Furthermore, as one of the most prominent features of active fluids, the emergent collective motion of self-driven units leads to swarming patterns \cite{Wensink12,Sokolov12}, which strongly affect systems' rotational degrees of freedom. Thus, the rotational dynamics of anisotropic particles also provide a way for probing intrinsic properties of the swarming phase of active fluids \cite{Sokolov10}. 

\begin{figure*}
\begin{center}
\includegraphics[width=7in]{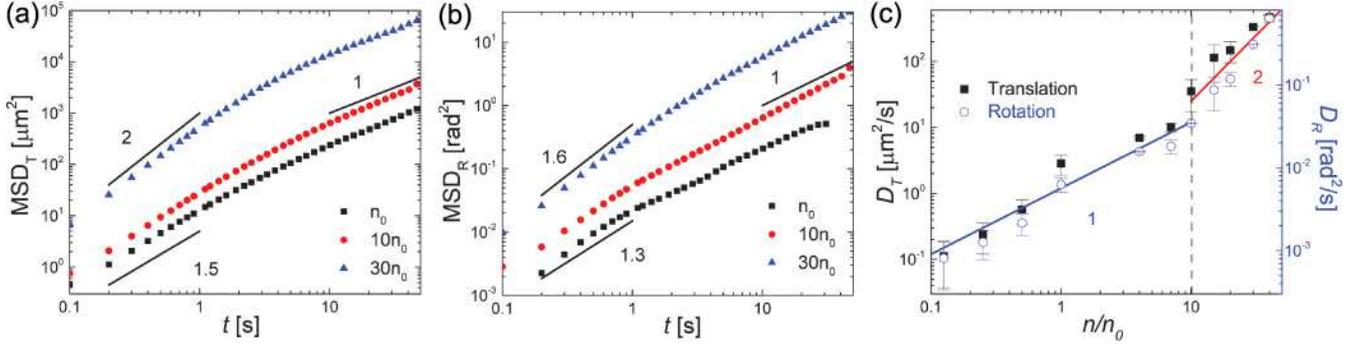}
\end{center}
\caption[Enhanced translational and rotational diffusions]{(Color online). Enhanced translational and rotational diffusions. (a) Translational mean-squared displacements (MSD$_\textrm{T}$) and (b) Rotational mean-squared displacements (MSD$_\textrm{R}$). The slopes of lines are indicated. (c) Translational and rotational diffusion coefficients, $D_T$ and $D_R$, as a function of bacterial concentrations, $n$. Solid lines indicate linear and nonlinear enhancements. The vertical line indicates the onset of obvious bacterial swarming.} \label{Figure2}
\end{figure*} 

In our experiments, we use suspensions of {\it Escherichia coli} as our model active fluids (see supplementary material \cite{supplementary}). A bacterial suspension of concentration $n$ is suspended in a free-standing soap film, where $n$ is measured in the unit of $n_0 = 8 \times 10^8$ cells/ml. The film is adjusted to be $15 \pm 5$ $\mu$m in thickness and $5 \times 5$ mm$^2$ in area. Polystyrene ellipsoids are used as our anisotropic tracer particles. In our study, we mainly focus on ellipsoids with semiprincipal axes of length $a = b = 2.8 \pm 0.2$ $\mu$m and $c = 14.2 \pm 0.5$ $\mu$m and aspect ratio $p \equiv c/a = 5.1$, although two other ellipsoids with different sizes and aspect ratios are also tested (see Ref.~\cite{supplementary} and Fig.~\ref{Figure4}).

Since the major axis of ellipsoids is larger than the thickness of film, particles undertake a quasi-two-dimensional motion in the bacterial bath. The diffusion of ellipsoids is recorded with high-speed optical microscopy (Fig.~\ref{Figure1}a), which allows us to extract the center-of-mass position, $\bm{x}(t) = [x(t),y(t)]$, and the orientation, $\theta(t)$, of ellipsoids at different times, $t$ (Fig.~\ref{Figure1}b). 

First, we analyze the translation and rotation of ellipsoids in the lab frame. Figures~\ref{Figure2}a,b show ellipsoids' translational mean-squared displacements (MSD$_\textrm{T}$), $\langle \lbrack \Delta \bm{x}(t) \rbrack \rangle^2 = \langle \lbrack \bm{x}(t+t_0)-\bm{x}(t_0) \rbrack^2 \rangle$, and rotational mean-squared displacements (MSD$_\textrm{R}$), $\langle \lbrack \Delta \theta(t)\rbrack^2\rangle = \langle \lbrack \theta(t+t_0) - \theta(t_0) \rbrack^2\rangle$, at different $n$, respectively.  The two degrees of freedom exhibit qualitatively similar trends. At short times, the particle motion appears to be super-diffusive following $\langle \lbrack \Delta \bm{x}(t) \rbrack^2 \rangle \sim t^{\delta_T}$ and $\langle \lbrack \Delta \theta(t) \rbrack^2 \rangle \sim t^{\delta_R}$ with the power exponents $1.5 \leq \delta_T \leq 2.0$ and $1.3 \leq \delta_R \leq 1.6$. At long times, the motion becomes diffusive with $\delta_{T,R} \approx 1.0$. The corresponding diffusion coefficients, defined in the diffusive regime as $D_T= \langle \lbrack \Delta \bm{x}(t) \rbrack^2 \rangle/4t$ and $D_R=\langle \lbrack \Delta \theta(t) \rbrack^2 \rangle/2t$, increase linearly with $n$ at low $n$ (Fig.~\ref{Figure2}c), similar to the translational diffusion of spherical particles in dilute suspensions of microswimmers \cite{Wu00,Chen07,Wilson11,Mino13,Jepson13,Valeriani11,Leptos09,Kurtuldu11,Kasyap14}. However, for $n \geq 10n_0$, where bacteria develop collective swarming motions (Fig.~\ref{Figure1}a), the enhancement of $D_{T,R}$ becomes nonlinear following a trend consistent with a power-law scaling $D_{T,R}\sim n^2$ (Fig.~\ref{Figure2}c).

\begin{figure*}
\begin{center}
\includegraphics[width=7in]{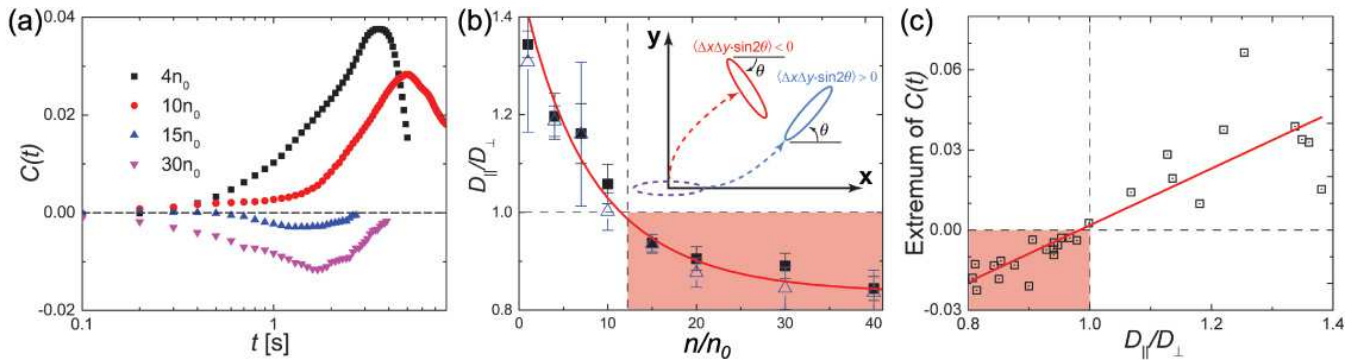}
\end{center}
\caption[Translation-rotation coupling]{(Color online). Translation-rotation coupling. (a) Cross correlations at different $n$. The dashed line indicates zero coupling. (b) The ratio of diffusion coefficients along the major and minor axes in the body frame. Solid squares are from direct measurements and blue triangles are the product of $\langle \Delta x_\parallel^2 \rangle/\langle \Delta x_\perp^2\rangle$ and $\tau_\parallel/\tau_\perp$ (see text). The solid line is an empirical relation for visual guide. Inset illustrates schematically the distinct particle motions with positive (blue) and negative (red) translation-rotation couplings. (c) The extremum of $C(t)$ v.s. $D_\parallel/D_\perp$. The solid line is a linear fit. The ``prohibited zone'' of Brownian motion is indicated by the red areas in (b) and (c).} \label{Figure3}
\end{figure*} 
  
The qualitative change of $D_{T,R}(n)$ scaling reflects the underlying transition of bacterial suspensions from the disordered to the swarming phase \cite{Wensink12,Sokolov12}. Generally, a diffusion coefficient, $D_{T,R} \sim l^2$, where $l$ is the step size of particle motion in a short time in the superdiffusive regime. $l$ is determined by the strength of the unbalanced fluid flow created by numerous swimming bacteria at the location of ellipsoids, which in turn is proportional to the fluctuation of bacterial concentration surrounding the ellipsoids, $\delta n$. Thus, $D_{T,R} \sim l^2 \sim \delta n^2$. The linear scaling $D_{T,R} \sim n$ at low $n$ indicates $\delta n \sim \sqrt{n}$, consistent with the central limit theorem \cite{Narayan07}. In contrast, the nonlinear enhancement $D_{T,R} \sim n^2$ at high $n$ indicates a giant number fluctuation $\delta n \sim n$, which is known as a hallmark of the swarming phase of active fluids \cite{Narayan07,Palacci13,Zhang10}.

\begin{figure}
\begin{center}
\includegraphics[width=2.5in]{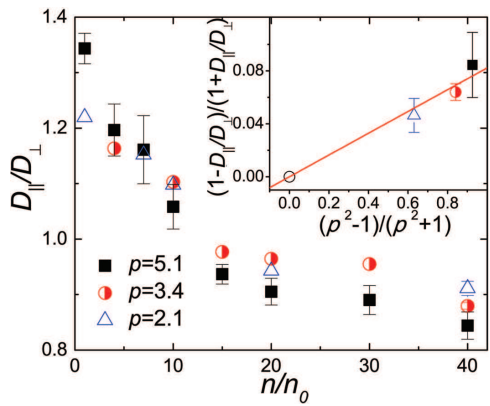}
\end{center}
\caption[aspect ratio dependence]{(Color online). $D_\parallel/D_\perp$ as a function of $n$ for ellipsoids of different aspect ratios, $p$. Inset shows the linear relation between $(1-D_\parallel/D_\perp)/(1+D_\parallel/D_\perp)$ and $(p^2-1)/(p^2+1)$, as predicted by the hydrodynamic model (Eq.~\ref{equ2}), where $D_\parallel/D_\perp$ are taken at $n = 40n_0$.} \label{Figure4}
\end{figure} 
   
Next, we investigate the coupling between the translational and rotational motions of ellipsoids by measuring a dimensionless cross-correlation function, $C(t)=\langle \Delta x \Delta y \sin 2\theta\rangle/\langle \lbrack \Delta \bm{x}(t) \rbrack^2 \rangle$ (Fig.~\ref{Figure3}a). $C(t)$ deviates from zero, revealing a coupling between the two degrees of freedom. More interestingly, $C(t)$ shows a transition from positive to negative correlations with increasing $n$. In sharp contrast, the translation and rotation of a Brownian ellipsoid are always positively coupled \cite{Han06}. The translational friction coefficient parallel to the major axis of an ellipsoid, $\zeta_{\parallel}$, is invariably smaller than that perpendicular to its major axis, $\zeta_{\perp}$---a generic feature of Stokes flow \cite{Happel83}. Thus, from the Einstein relation, a Brownian ellipsoid diffuses faster along its major axis with $D_{\parallel}/D_{\perp}>1$, where $D_i = k_BT/\zeta_i$ is the diffusion coefficient in the body frame of the particle with $i =$ $\parallel$ (parallel to the major axis) or $\perp$ (perpendicular to the major axis) and $k_BT$ is the thermal energy. Hence, an ellipsoid undertaking Brownian motion has a tendency to diffuse along its major axis, resulting in a positive translation-rotation coupling in the lab frame as shown schematically by the blue ellipsoid in the inset of Fig.~\ref{Figure3}b. Note that when diffusing faster along its major axis, the ellipsoid rotates counterclockwise, leading to a positive $\theta$ and, therefore, $\langle \Delta x \Delta y \sin 2\theta\rangle>0$. In contrast, the negative coupling indicates $D_{\parallel}/D_{\perp}<1$ (the red ellipsoid in Fig.~\ref{Figure3}b inset), a counterintuitive result strictly prohibited in Brownian diffusion. Thus, anisotropic particles in bacterial suspensions possess an unusual mode of translation-rotation coupling, non-existent in equilibrium systems.
 
Motivated by the above analysis, we directly measure anisotropic particle diffusions in the particles' body frame. We transform the trajectories of ellipsoids into the body frame and obtain MSDs parallel and perpendicular to their major axes \cite{supplementary}. The body-frame MSDs show the same transition from the super-diffusive to the diffusive behaviors, allowing us to extract $D_i$ at long times. Consistent with the analysis of the lab-frame correlations, $D_\parallel/D_\perp$ decreases with $n$ and enters into ``the prohibited zone'' of Brownian diffusion at high $n$ with $D_\parallel/D_\perp<1$ (Fig.~\ref{Figure3}b). The link between the coupling mode and the anisotropic body-frame diffusion is further illustrated by the linear relation between $D_\parallel/D_\perp$ and the extremum of $C(t)$, where the emergence of negative $C(t)$ coincides with $D_\parallel/D_\perp<1$ (Fig.~\ref{Figure3}c). $D_\parallel/D_\perp$ shows a qualitatively similar trend for ellipsoids of other aspect ratios, $p$ (Fig.~\ref{Figure4}). But the variation of $D_\parallel/D_\perp$ with $n$ reduces with decreasing $p$. 

\begin{figure*}
\begin{center}
\includegraphics[width=7in]{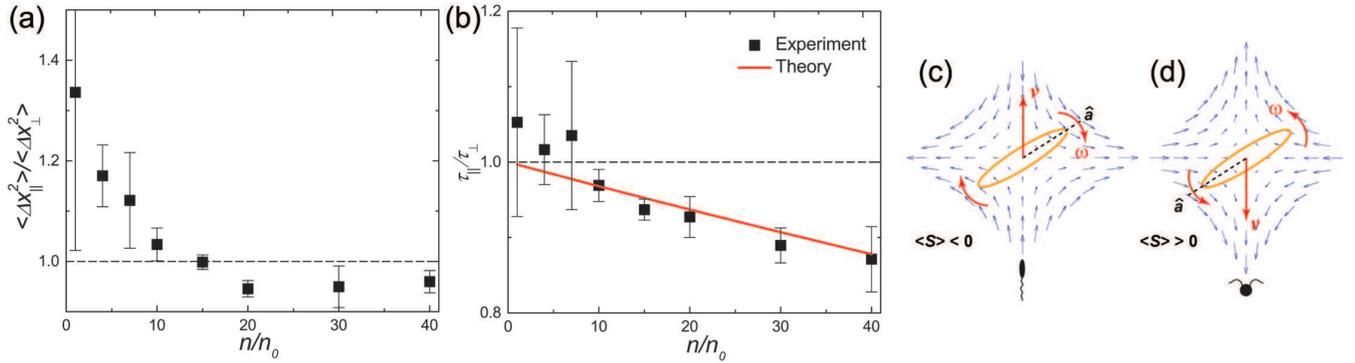}
\end{center}
\caption[Origin of the anomalous coupling]{(Color online). Origin of the anomalous translation-rotation coupling. (a) Ratio of short-time MSDs along the major and minor axes in the super-diffusive regime with $\Delta t = 0.1$ s. (b) Ratio of correlation times along the major and minor axes. Schematics showing the average effective straining on ellipsoids induced by the extensile dipole flow of pushers such as {\it E. coli} (c) and the contractile dipole flow of pullers such as {\it C. reinhardtii} (d).} \label{Figure5}
\end{figure*}

To understand the origin of the anomalous anisotropic diffusions, we apply the Green-Kubo formula \cite{Zwanzig01}, where $D_i$ is expressed as the integral of the velocity auto-correlation, $D_i= \int_0^\infty dt \langle v_i(t+t_0)v_i(t_0)\rangle$, which is further approximated as $D_i \approx \langle v_i^2 \rangle \int_0^\infty dt \langle n_i(t+t_0)n_i(t_0)\rangle$. Here, $n_i(t) \equiv v_i(t)/|v_i(t)|$  is the direction of particle velocity $v_i$ along the particle's major ($i=$ $\parallel$) or minor ($i=$ $\perp$) axis at $t$. Thus, $D_i$ is decomposed into two parts: (1) the average particle velocity in the super-diffusive regime quantified by MSDs in a short-time interval $\Delta t$, $\langle v_i^2 \rangle = \langle \Delta x_i^2 \rangle/\Delta t^2$; and (2) the auto-correlation of the velocity direction, $A_i(t)\equiv\langle n_i(t+t_0)n_i(t_0)\rangle$. $A_i(t)$ decays exponentially following $A_i(t)=\exp(-t/\tau_i)$ \cite{supplementary}, where the correlation time, $\tau_i$, indicates the persistence of the motion. Hence, $D_i=\langle v_i^2 \rangle \tau_i$, leading to $D_\parallel/D_\perp = (\langle \Delta x_\parallel ^2 \rangle/\langle \Delta x_\perp^2 \rangle)(\tau_\parallel/\tau_\perp)$ (Fig.~\ref{Figure3}b). At low $n$, while $\tau_\parallel$ and $\tau_\perp$ are comparable (Fig.~\ref{Figure5}b), $\langle \Delta x_\parallel ^2 \rangle$ is consistently larger than $\langle \Delta x_\perp ^2 \rangle$ (Fig.~\ref{Figure5}a), a feature similar to Brownian motion of anisotropic particles \cite{Han06,Mukhija07}. However, as $n$ increases, $\langle \Delta x_\parallel ^2 \rangle / \langle \Delta x_\perp ^2 \rangle$ approaches to a constant close to 1, whereas $\tau_\perp$ grows larger than $\tau_\parallel$, resulting in the anomalous $D_\parallel/D_\perp < 1$. Thus, the anomalous coupling arises from the more persistent motion of an anisotropic particle along its minor axis.

To quantitatively interpret the anomalous diffusions from a first-principles calculation, one needs to consider hydrodynamic interactions among a large number of swarming bacteria, which is notoriously complicated and is still far from well understood \cite{Koch11,Marchetti13,Poon13}. Such a theoretical investigation is out of the scope of our current experimental study. Nevertheless, we find that the decreasing trend of $\tau_\parallel/\tau_\perp$ emerges even when only considering the far-field dipole flow of a single bacterium (see SI for detailed discussions \cite{supplementary}). 

In short, we extend previous models on enhanced diffusion of point-like spherical particles \cite{Underhill08,Ishikawa10,Lin11,Pushkin13,Morozov14} by assuming point-like particles with anisotropic shapes. The intrinsic correlation between the translation and rotation of an anisotropic point particle can be quantified using a scalar $S \equiv \bm{\omega} \cdot \left( \frac{\bm{\hat{a}} \times \bm{v}}{|\bm{\hat{a}} \times \bm{v}|}\right)$, where $\bm{v}$ and $\bm{\omega}$ are the translational and angular velocity of the particle in a dipole flow created by a swimming bacterium \cite{Drescher11}, and $\bm{\hat{a}}$ is the unit vector along the major axis of the particle that forms an acute angle with $\bm{v}$ (Figs.~\ref{Figure5}c, d). We demonstrate that the spatial average of $S$, $\langle S \rangle$, is negative for the extensile dipole flow of {\it E. coli} \cite{supplementary}. The negative sign indicates that the particle rotates in a direction that aligns its minor axis with the direction of its own translation, similar to the motion of a particle in a straining flow (Fig.~\ref{Figure5}c), whereas the magnitude, $|\langle S \rangle|$, gives the average speed of such a rotation. Under the influence of random fluctuations in a bacterial bath, this effective straining leads to \cite{supplementary}:
\begin{equation}
\label{equ1} {\frac{\tau_\parallel}{\tau_\perp} = \frac{1+2\langle S \rangle t_c/\pi}{1-2\langle S \rangle t_c/\pi}},
\end{equation}   
where the dimensionless quantity, $2|\langle S \rangle|t_c/\pi$, is the ratio of the correlation time of random fluctuations \cite{Wu00}, $t_c$, to the characteristic time of the straining-induced rotation, $(\pi/2)/|\langle S \rangle|$. Eq.~\ref{equ1} successfully predicts $\tau_\parallel / \tau_\perp < 1$  for $\langle S \rangle < 0$. Moreover, $2|\langle S \rangle|t_c/\pi$ should be zero at $n = 0$ and increase with $n$. Indeed, a linear approximation, $2|\langle S \rangle|t_c/\pi = c(n/n_0)$, quantitatively fits our experiments with a numerical constant $c = 1.6 \times 10^{-3}$ (Fig.~\ref{Figure5}b). Although the assumption of our current analysis only validates at low $n$, where the far-field dipole flow dominates the bacteria-particle interaction, the quantitative agreement between experiments and this linear approximation indicates that at the coarse-grained level the swarming of a group of bacteria may still be treated as an effective dipole, albeit with a greatly enhanced dipolar strength. Last, the simple hydrodynamic model also predicts that the strength of the local straining decreases with the ellipsoids' aspect ratio, $p$, and vanishes for spherical particles with $p = 1$, which, at high $n$, leads to a relation \cite{supplementary}:
\begin{equation}
\label{equ2} {\frac{1-(\tau_\parallel/\tau_\perp)}{1+(\tau_\parallel/\tau_\perp)} \approx \frac{1-(D_\parallel/D_\perp)}{1+(D_\parallel/D_\perp)} \sim \frac{p^2-1}{p^2+1}},
\end{equation}  
consistent with our experiments (Fig.~\ref{Figure4} inset).     
 
A self-propelled microswimmer can be generally categorized as either a ``pusher'' or a ``puller'' depending on the far-field flow it creates \cite{Koch11,Marchetti13,Poon13}. Our simple calculation showed that the anomalous translation-rotation coupling is a consequence of extensile dipole flows---a defining feature of pushers. Thus, this coupling should be universal for all pusher-type active fluids. In contrast, for puller microswimmers such as {\it Chlamydomonas reinhardtii}, $\langle S \rangle$ is positive (Fig.~\ref{Figure5}d) \cite{supplementary}. Therefore, an anisotropic particle in puller-type of active fluids should exhibit an enhanced {\it positive} translation-rotation coupling. However, due to the lack of the swarming behavior of pullers \cite{Koch11}, the effect is likely to be weak. Our preliminary experiments with {\it C. reinhardtii} are consistent with this prediction. 

In addition to providing a new insight into the intrinsic properties of active fluids, our results also deepen the understanding of transport processes in microbiological systems \cite{Berg93}. We have demonstrated that the translation and rotation of anisotropic particles---naturally the most abundant particles in microbiological systems (e.g. dead bacterial bodies and macromolecules secreted by bacteria)---are not two {\it decoupled} degrees of freedom. Instead, the rotation of an anisotropic particle profoundly influences how it translates and thus fundamentally modifies its transport dynamics.       

\begin{acknowledgments}
We thank K. Dorfman for providing us with the fluorescently tagged {\it E. coli} strain. We also thank O. Yang, C. Tang and P. Lefebvre for the help with experiments on {\it C. reinhardtii}. The research is partially supported by ACS Petroleum Research Fund No. 54168-DNI9. L.L. and X.X. acknowledge the support by National Natural Science Foundation of China No. 11575020.
\end{acknowledgments}

\end{document}